\documentclass[doublecol]{epl2}
\usepackage{latexsym}
\title{Nonlocal Granular Rheology: Role of Pressure and Anisotropy}
\shorttitle{Nonlocal Granular Rheology: Role of Pressure and Anisotropy} 

\author{Elie Wandersman\inst{1,2} \and Martin van Hecke\inst{1}}
\shortauthor{E. Wandersman and M. van Hecke}

\institute{
  \inst{1} - Kamerlingh Onnes Laboratory, Leiden University, PObox 9506, 2300RA, Leiden, The Netherlands \\
  \inst{2} - Laboratoire Jean Perrin, FRE 3231, Universit\'e Pierre et Marie Curie - CNRS, 4, Place Jussieu, 75005 Paris -- France
}

\abstract{We probe the secondary rheology of granular media, by imposing a main
flow and immersing a vane-shaped probe into the slowly flowing granulate. The secondary  rheology is then the relation between the exerted torque $T$ and rotation rate $\omega$ of our probe. In the absence of any main flow, the probe experiences a clear yield-stress, whereas for any finite flow rate,
the yield stress disappears and the secondary rheology takes on the form of a double exponential relation between $\omega$ and $T$. This secondary rheology does not only depend on the magnitude of $T$, but is anisotropic --- which we show by varying the relative orientation of the probe and main flow. By studying the depth dependence 
of the three characteristic torques that characterize the secondary rheology, we show that for counter flow, the dominant contribution is frictional like --- i.e., $T$ and pressure are proportional for given $\omega$ --- whereas for co flow, the situation is more complex. Our experiments thus reveal the crucial role of anisotropy for the rheology of granular media.
}

\begin{document}

\maketitle

We still lack a full description of slow, dense flows of granular media --- given boundary conditions, what determines the flow rate? Granular flows differ in two important ways from Newtonian flows. First,
friction plays a central role --- when grains have persistent contacts, as is the case in slow granular flows, inter-particle friction provides the main channel by which energy can be dissipated. Experiments in which the resistance to granular flow is probed by controlling the stress on a single moving boundary, such as in Couette, split-bottom or vane geometries \cite{DijksmanPRL11,BehringerNature03,DijksmanPRE10}, find that the shear stresses are proportional to the confining pressure, and that the ratio of shear to compressive stress, which can be seen as an effective friction coefficient, does not vary strongly with rate \cite{PouliquenAnnRevFlu08,GDRMidi,BehringerNature03,DijksmanPRL11, DijksmanPRE10}. The crucial consequence from this rate independence is that the stress plateaus for strain rates going to zero.
In other words, the stresses in slow granular flows are rate independent, and therefore the stress is not sufficient to set the strain rate.

The second ingredient is non-locality. Several recent experiments \cite{BocquetNature08,NicholPRL10,PouliquenPRL11} and theoretical works \cite{KamrinPRL12,AndreottiArxiv13,BocquetPRL09,KamrinPNAS13}
indicate that for matter with granularity, the ''fluidity'' in location A, i.e., the local relation between stress and strain rate, can be strongly influenced by the flow in location B. Such non-local behavior was first observed in the flow of emulsions \cite{BocquetNature08}, has also
been observed in foams \cite{KatgertEPL10}, and has been modeled by
a diffusive model for the fluidity, leading to the introduction of a length scale which characterizes the nonlocality \cite{BocquetNature08,AndreottiArxiv13,BocquetPRL09}.  Kamrin and coworkers have adapted these ideas
for frictional rheologies, capturing granular Couette flows \cite{KamrinPRL12}, and more recently, the full flow profiles and height dependence of split bottom granular flows \cite{KamrinPNAS13,FenisteinNature04,FenisteinPRL04,FenisteinPRL06,ChicagoPRL06,joshuareview}.

In all these studies, there is a single source driving the flow, and the nonlocality manifests itself via the spatial flow profiles. However, as flow in location B influences the fluidity in A, it is natural to study situations where the driving of the main flow and the probing of the rheology are independent. Two examples of such granular experiments which probe the ``secondary rheology'' concern the sinking of a passive probe into a granular bed that is stirred far away from the probe (in a split-bottom cell) \cite{NicholPRL10}, and the rheology probed by rods submersed in and dragged through
a granular medium that is itself stirred in a Couette geometry \cite{PouliquenPRL11}.
Recently, numerical studies in which a plate is dragged through a 2D granular simple shear flow have been performed \cite{AndreottiArxiv13}.
In all cases, the yield stress experienced by the probes was found to vanish as soon as there is any external flow imposed, and the rate of the probe was proportional to the stirring rate, supporting the qualitative picture, that flow in one location leads to mechanical agitations that govern the rheology in locations further away. Open questions concern: {\em(1)} the nature of the local rheology under stirring --- in \cite{NicholPRL10} the relation between stress and strain rate appears linear, whereas in \cite{PouliquenPRL11,AndreottiArxiv13} it appears exponential; {\em(2)} the effect of pressure --- are shear stresses still proportional to pressure? and {\em(3)} the role of anisotropy --- granular flows lead to anisotropy of the granular fabric \cite{LosertPRL04,BehringerNature05} which should influence the fluidity of the material.

\begin{figure}[tbp]
\begin{center}
\includegraphics[width=1\linewidth,clip,viewport=0 150 580 420]{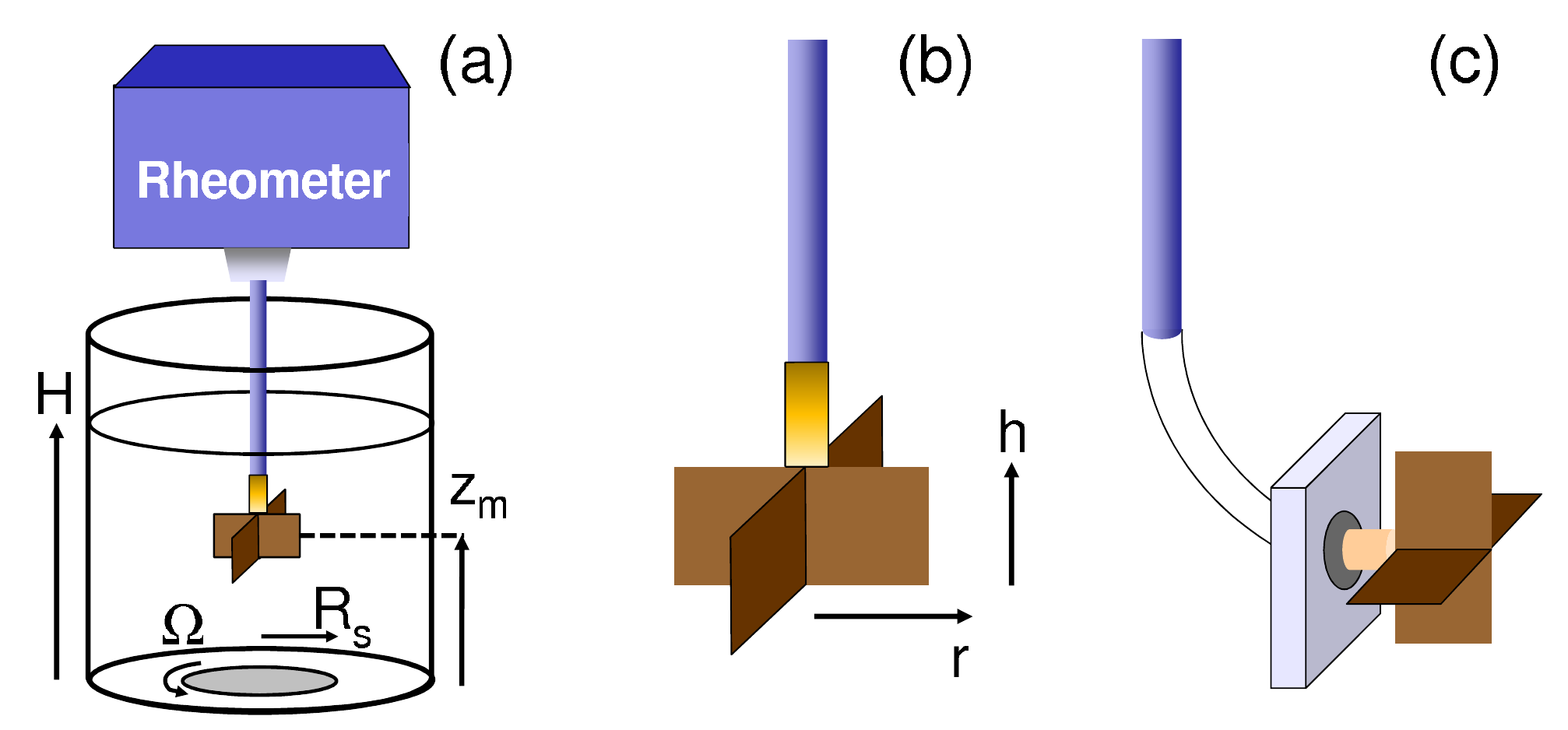}
\end{center}
\caption{(Color online) (a) Experimental setup. A disk of radius $R_s$ rotating at a rate $\Omega$
in a split-bottom flow cell filled to height $H$ with 1 mm glass beads causes the main flow. A vane of
radius $r \ll R_s$ and height $h$, connected to a rheometer, probes the secondary rheology at a depth $z_m$.
(b) Most of our experiments are carried out with a vane with its central axis identical to the central axis of the rotating disk. (c) We also performed experiments with a vane with its central axis parallel to the bottom disk, so that the secondary flow induced by the probe is perpendicular to the main flow. }\label{f1}
\end{figure}

Here we probe these questions by characterizing the secondary rheology in a split bottom cell where the main flow is driven independently (Fig.~\ref{f1}). Our experiments allow us to determine full secondary rheological curves for a range of driving conditions and pressures. Moreover, we can probe the rheology when the probe moves with the main flow (coflow), against the main flow (counterflow) and
also for probe rotation perpendicular to the main flow (Fig.~\ref{f1}b-c).

We find that, consistent with earlier work, that whenever the main flow is present, the secondary rheology has a vanishing yield threshold \cite{NicholPRL10,PouliquenPRL11,AndreottiArxiv13}. The local rheology is anisotropic, and there are significant qualitative differences between co flow, counter flow, and perpendicular flow. Finally, by probing the flow at different depths, we find that the characteristic shear stresses are in some, but not all cases, proportional to the compressive stresses. Our results show that both local pressure and local anisotropy are crucial to describe the fluidity of granular media, thus stressing that these aspects need to be incorporated in future theories of slow granular flows.

{\em Setup and protocol ---} We create a smooth granular flow in a
split-bottom shear cell \cite{FenisteinNature04,FenisteinPRL04,FenisteinPRL06,ChicagoPRL06,joshuareview,NicholPRL10}
as shown in Fig.~\ref{f1}a. This cell consists of
a cylindrical container of radius 80 mm, the bottom of which is split into a rotating disk of radius $R_s = 60$ mm and a stationary outer ring. The cell is made of aluminum and the bottom and side walls are made rough by drilling dimples in them so as to
obtain granular no-slip boundary conditions. The disk is driven by a micro-stepping motor at angular velocities $\Omega$ ranging from $10^{-3}$ to $1$ rps.
The split-bottom cell is filled with a controlled mass of 1 mm diameter glass beads up to the filling heights $H$. We focus on large filling heights (
$H$= 60  mm), as the flow then predominantly takes place in a dome-shaped
shear band buried in the granular medium \cite{FenisteinPRL04,FenisteinPRL06,ChicagoPRL06,UngerPRL04} --- this is the same flow geometry as used in \cite{NicholPRL10}.

To measure the rheological properties of this fluidized medium,
we immerse a vane-shaped intruder into the material and couple it to an Anton Paar DSR301 rheometer. As illustrated in Fig.~\ref{f1}b,c, the vane is made of crossed rectangular metallic plates of thickness 1 mm and of radius $r$ by height $h$
5x5, 10x10 or 20x20 mm. In most experiments, this probe is directly attached to the rheometer via a smooth and rigid shaft (Fig.~\ref{f1}b),
and  we have checked that the residual torques picked up by the shaft are
negligible in comparison to the signal from the vane. We have also carried
out some experiments where we rotate the orientation of the vane as shown in Fig.~\ref{f1}c, and where we use a flexible tube with sufficiently large torsional stiffness to couple the vertical rheometer shaft to the horizontal vane shaft. The vertical distance between the center of the probe and the free surface of the material, $z_m$ can be adjusted.

The rheometers native operation is in torque controlled mode, where torques
ranging from -20 to 20 mNm with a precision better than 1 $\mu$Nm are applied to our vane and the resulting deflection angle of the probe $\phi(t)$ is measured. The precision on $\phi$ is estimated to be significantly better than $10^{-4} rad$, and in all experiments we deduce the average rotation rate $\omega := \partial_t \phi(t)$ over
deflection angles of 0.6 rad or more. We define negative (positive) values of $\omega$ to correspond to counter-rotation (co-rotation) with the bottom disk.
Transients are short and all data presented
are steady state values.

Prior to each measurement, the beads are vigorously stirred with a rod, after which the probe is immersed to its position $z_m$. The system is then pre-sheared with $\Omega=0.5$ s$^{-1}$ for 10 minutes, while the humidity of the system is lowered to values of 10 - 20 \% relative humidity by closing the system with a plastic cover and injecting a dry air flow. This ensures a very good reproducibility of our rheological measurements.

{\em Precession and Yielding ---} Ultimately, our goal is to probe and understand
the relation between the probe rotation $\omega$, disk rate $\Omega$ and driving torque $T$,
$\omega(\Omega,T)$, for a range of depths $z_m$ and dimensions of
the probe. Before delving into the full rheology, we first focus on the two single-forcing regimes: first
 the passive
rotation of the probe due to residual flow in the system, $\omega(\Omega,T=0)$, and second  the rheology of the sand in absence of stirring, $\omega(\Omega=0,T)$.

\begin{figure}[tbp]
  \centering
  \includegraphics[width=\columnwidth,clip,viewport=20 280 550 520]{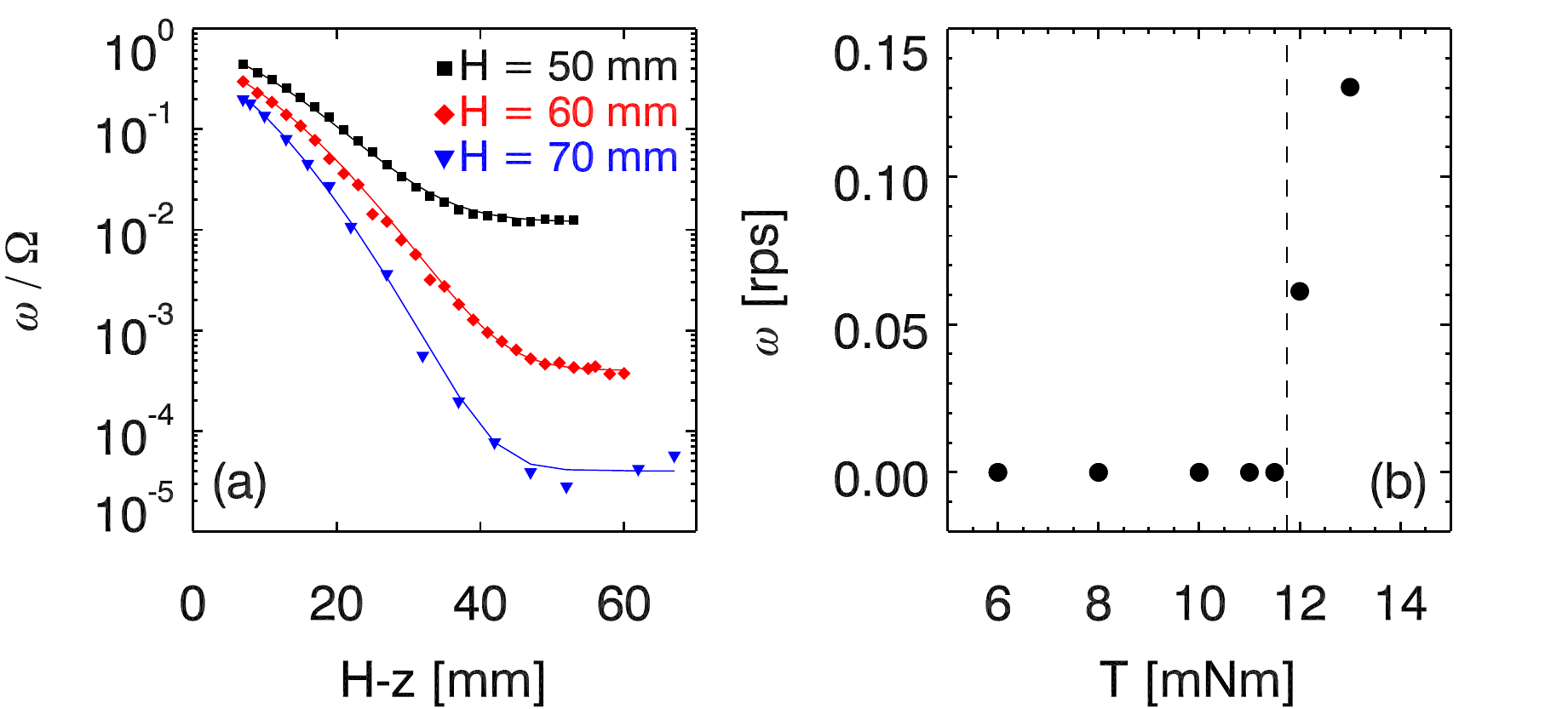}
  \caption{(Color online) (a) The residual rotation rate $\omega_p$ of the probe in absence of any torque on the probe is plotted as a function of depth $H-Z$ --- curves are stretched exponential fits given by Eq.~(1). (b) Rotation rate $\omega$ of a $20\times 20$ mm vane in absence of a base flow ($\Omega=0$) as  function of driving torque indicates a yield torque of $11.7 \pm 0.3$ mNm.}\label{f2}\end{figure}

We show in Fig.~\ref{f2}a the behavior in the absense of a driving torque. As is well known, the rotation of the driving disk sets up a residual torsional flow in the medium \cite{ChicagoPRL06,FenisteinPRL06,NicholPRL10}, that we
characterize here by immersing one of our probes in the medium at depth $z_m$, setting the torque equal to zero, and measuring the ensuing precession of our probe, $\omega_p(z)$. For the three filling heights investigated, we find that $\omega_p(z)$ decreases with $z$ as a stretched exponential:
\begin{equation}
\omega_p(z)/\Omega = A e^{-(\frac{H-z}{\xi})^\alpha} +B~,
\end{equation}
where $A$ and $B$ are determined by the precession rates at $z=0$ ($\omega_p$) and
at $H-z=0$ \cite{NicholPRL10} --- $\alpha \approx$ 1.5 and $\xi$ represents a characteristic size of the main flow and is of the order of ten grain diameters. Moreover, the precession velocity is found to be independent of the probe size (not shown), and proportional to $\Omega$.

We now turn to the rheology of our immersed probes for $T\neq 0$. We first
establish that for $\Omega =0$, we recover the typical yield stress behavior of
granular media. As indicated in Fig.~2b, experiments where we slowly ramp up $T$ then reveal a yield torque $T_y = 11.7 \pm 0.3$ mNm, above which there is rapid flow, and below which there is, in essence, no flow \footnote{If we quench $T$ from zero to a value close to $T_y$, we can detect a very slow creep flow that has a rate that decreases in time and tends to zero.}

To extract the static friction coefficient $\mu_s$ from the observed yield torque, we make the following three standard assumptions for frictional granular media:
{\em(i)} The pressure is hydrostatic: $P= \rho g z$, where for our material $\rho=1850$ kg/m$^3$; {\em(ii)} The shear stresses are maximal, and the slip flow localizes at
the cylindrical hull of the probe (area $2\pi r_e h_e$), where $r_e$ and $h_e$ are the effective dimensions of the probe ---
from earlier work, we know we can estimate $r_e$ and $h_e$ by extending the bare probe dimensions by halve a grain diameter,
i.e., $r_e=r+0.5$ mm and $h_e=h+1$ mm
\cite{NicholPRL10}; {\em(iii)}
The yielding shear stress is proportional to the product of friction coefficient and pressure:
$\sigma_{r\theta} (z)= \mu_s P(z) = \mu \rho g z$.
We find that
\begin{eqnarray}\label{lambdaeq}
T=r \int_A dA \sigma_{r \theta} = 2\pi r_e^2 \mu \rho g z_m h_e \Rightarrow \\
T/\lambda = \mu \rho g z_m = \mu P, \mbox{ where } \lambda:=  2 \pi r_e^2 h_e~.
\end{eqnarray}
Here $\lambda$ collects all geometric factors needed to relate torques to stresses (when the probe is
not fully submerged, we need to adapt $h_e$).
Substituting the relevant quantities for the experiments shown in Fig.~2, $r_e=20.5$ mm,
$h_e=21$ mm, and $z_m=20$ mm, we estimate that $\mu=0.58 \pm 0.1$ --- consistent with independent measurements of $\mu$ for these beads which are in the range of 0.52-0.58 \cite{DijksmanPRE10,DijksmanPRL11}.

{\em Secondary Rheology ---} Now we turn our attention to the secondary rheology when both the imposed flow rate $\propto \Omega$ and the driving torque of the vane, $T$, are non zero. Experimentally, we typically control $\Omega$ and $T$ and then measure the rotation rate $\omega$. We found that after a short transient, the rotation rate of the probe $\omega$ quickly reaches a steady state, and will show below that experiments where we control $\omega$ and measure $T$ yield identical results. We have also checked that runs done by sweeping the torque up or down are identical, and we have seen no history dependence, ruling out
thixotropic or other more complex rheologies. Hence, $\omega(T)$ is a well defined function fully determining the steady-state secondary rheology once the experimental parameters $\Omega$, $z_m$, $H$ and probe size are set.

\begin{figure}[tbp]
\includegraphics[width=\columnwidth,clip,viewport=15 120 376 400]{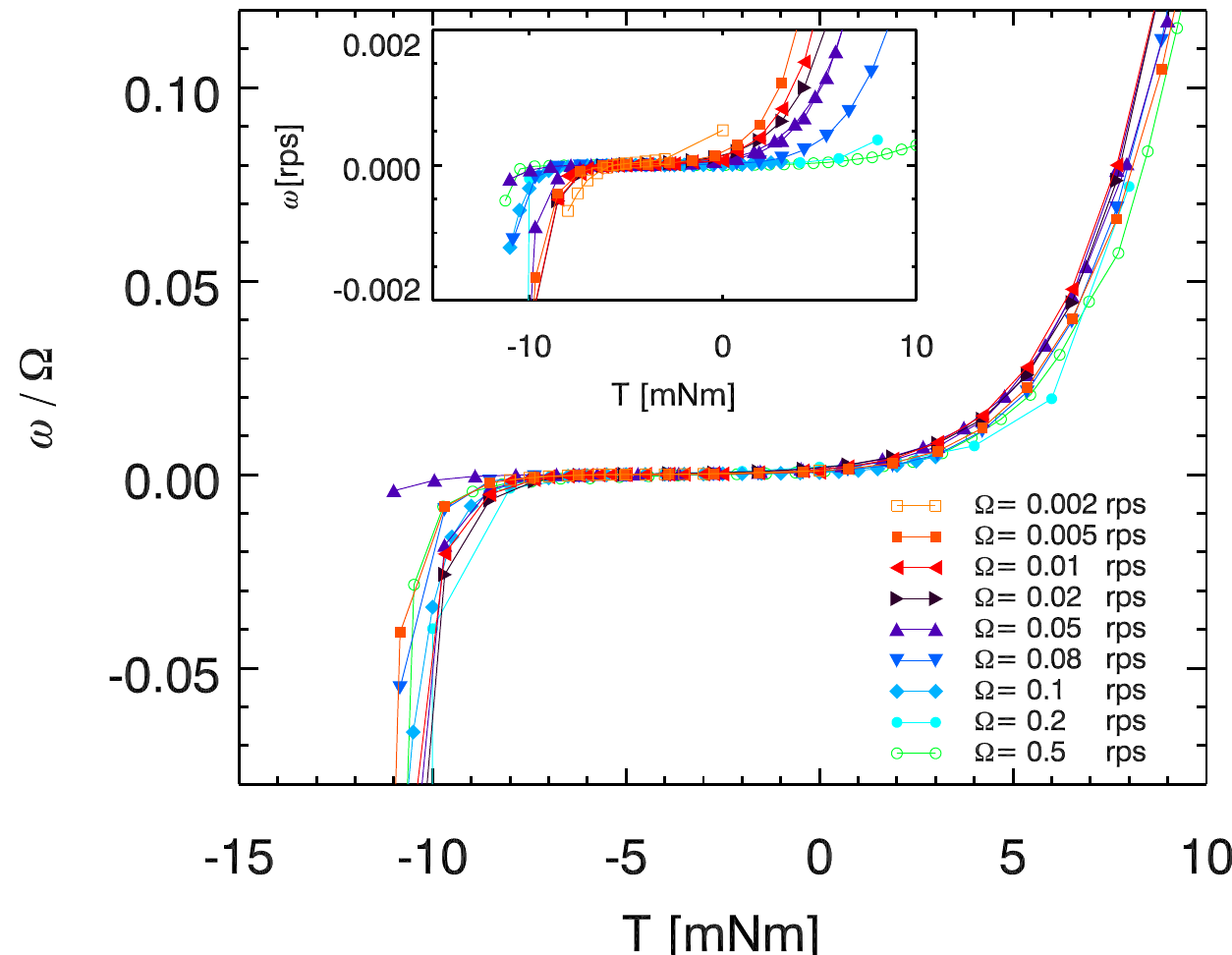}
\caption{(Color online) (a) Flow curves $\omega/\Omega$ for $H=60$ mm, for a 20x20 mm probe at $z_m=20$ mm, for a range of driving rates $\Omega$ as indicated. Inset: unscaled data  $\omega(T)$.
}\label{f3}
\end{figure}

In Fig.~\ref{f3} we show examples of such a rheological curve, $\omega(T)$,
for a $20x20$ mm probe buried at
depth $z_m=20$ mm, filling height $H=60$ mm and a range of rotation rates $\Omega$.
As expected \cite{FenisteinPRL04,FenisteinPRL06,ChicagoPRL06,NicholPRL10,PouliquenPRL11}, the overall driving rate $\Omega$ sets the only relevant timescale, so that $\omega \propto \Omega$, and all rheological curves collapse when plotted as $\omega/\Omega$ vs $T$.

\begin{figure*}[tb]
\includegraphics[width=2\columnwidth,clip,viewport=00 50 570 240]{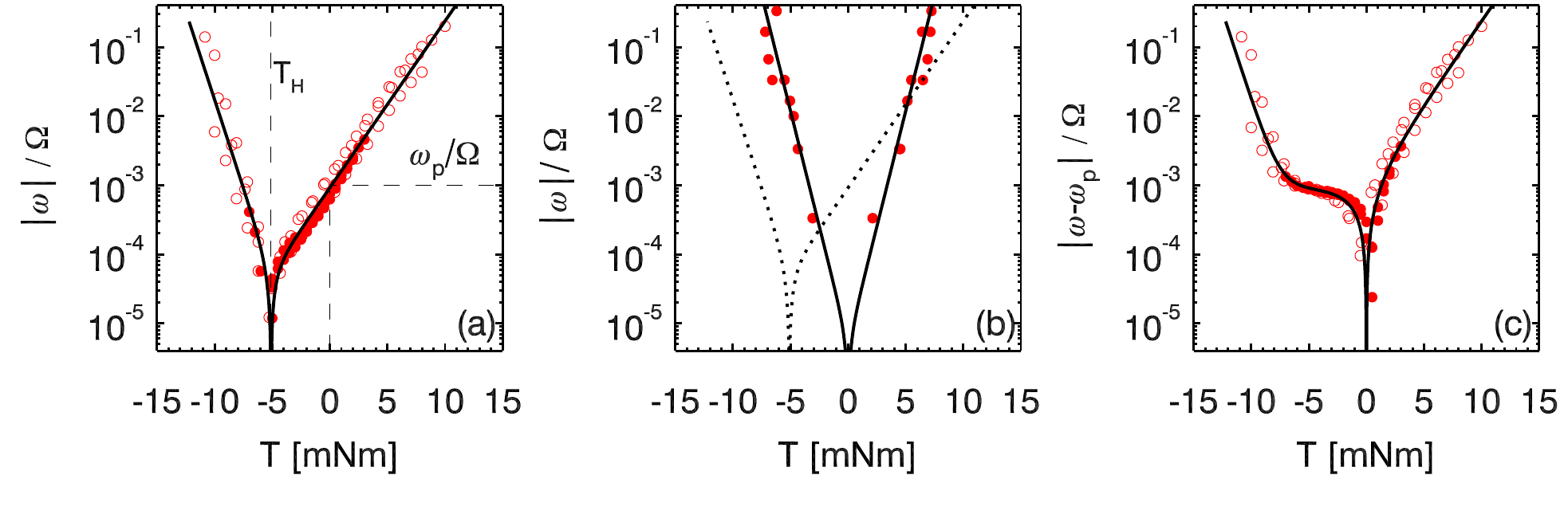}
\caption{(Color online) (a)
Rheological curves obtained for $H=60$ mm, for a $20\times 20$ mm probe, $z_m=20$ mm
and $\Omega$=0.05 rps. Full symbols corresponds to torque imposed experiments, open symbols to $\omega$ imposed experiments. The line is a  fit to the data using Eq.~(\ref{rheofit}) with $T_H=-5.1$ mNm, $T_- = 0.85 $ mNm and
$T_+ = 1.8 $ mNm.
(b) Symbols: Rheological curves for the same experimental parameters, but now for a probe in the perpendicular orientation such as shown in Fig.~1c. The full line is a fit to Eq.~(\ref{rheofit}) with $T_H=0$ mNm, $T_- = T_+ = 0.65 $ mNm; the dashed line is the fit from panel (a) shown for comparison. (c) If we replot the data shown in panel (a) but first subtract the local precession rate, the resulting curve does not exhibit a simple form. }\label{f4}
\end{figure*}

Our data shows that there is a clear difference between co flow ($\omega/\Omega >0$) and
counter flow ($\omega/\Omega <0$), and moreover suggests that for any finite value of $\Omega$, the yield stress vanishes so that there is no extended range of torques where the probes rotation rate $\omega=0$. To clarify these points, we present in Fig.~\ref{f4}
a plot of $|\omega|/\Omega$ for $\Omega=0.05 $ rps on a log scale. Fig.~\ref{f4}a confirms that there indeed is not finite yielding threshold, as there is no extended ``dead zone'' where $\omega=0$. Note that the precession flow induced by the rotation of the driving disk (Fig.~\ref{f2}a) means that for $T=0$, the vane rotation rate is equal to the precession rate, and indeed, for our data shown in Fig.~\ref{f4}a, we find that $\omega/\Omega$ is of order $10^{-3}$ for $T=0$, consistent with Fig.~2a. To hold the vane completely fixed in the lab frame, we need to apply a negative ''holding'' torque $T_H \approx -5.1$ mNm. In Fig.~\ref{f4}a we also show
that data obtained by torque control, i.e., setting $T$ and measuring $\omega$ (full symbols), and by rate-control, i.e., setting $\omega$ and measuring $T$ (open symbols), are identical within the experimental reproducibility.

Fig.~\ref{f4}a reveals that in very good approximation, $\omega$ varies as the sum of two exponential functions of $T$, and
all our data can be fitted extremely well by the following form:
\begin{equation}
\label{rheofit}
\omega/\Omega=C\left(e^{\frac{T-T_H}{T_+}}-e^{-\frac{T-T_H}{T_-}}\right)~.
\end{equation}
Here $T_H$ is the holding torque, and $T_+$ and $T_1$ determine the exponential growth of $\omega$ with torque for co- and counterflow. We note that the prefactor $C$ is directly related to physical parameters
$\omega_p$, $T_H$, $T_+$ and $T_-$; by substituting $T=0$, we immediately find
\begin{equation}
C =  \frac{\omega_
p/\Omega}{(e^{-T_H/T_+}-e^{T_H/T_-})}~.
\end{equation}

{\em Symmetry breaking ---} Our rheological data clearly shows that there
is a significant difference between co and counter rotation of the probe and disk.
This difference is easy to understand qualitatively, as the rotation of the bottom disk breaks the chiral symmetry.
The 3D nature of our flow allows us to probe the rheology in cases when the induced flow by the probe is neither opposite nor parallel to the disk-induced anisotropy, by using a ``perpendicular'' orientation of the probe as indicated in Fig.~\ref{f1}c. As shown in Fig.~\ref{f4}b, there is now
no symmetry breaking between left and right handed rotation of the probe. In good approximation, the rotation is still exponential in $T$, and a fit
to Eq.~(\ref{rheofit}) yields that for this case $T_H\approx 0 $ mNm, whereas $T_+ \approx T_- \approx 0.65$ mNm.

We now return to the broken symmetry case. How to understand the difference between co and counter flow? A subtlety in our experiment is that a priori it is not clear whether co and counter rotation should be defined with respect to the global disk rotation $\Omega$ (i.e., $\omega<0$ or $\omega>0$) or the local precession $\omega_p$ (i.e., $\omega<\omega_p$ or $\omega>\omega_p$). Our fitting form, Eq.(\ref{rheofit}), tacitly assumes that co and counter rotation should be defined with respect to the disk rotation, $\Omega$.
Alternatively, we could first correct for the local precession $\omega_p$, but
as Fig.~\ref{f4}c shows, deining co and counter flow with respect to the precession rate $\omega_p$, by plotting $(\omega-\omega_p)/\Omega$, hides the simple torque dependence shown in Fig.~\ref{f4}a. A crucial aspect of Fig.~4a is thus that it shows that when the torque is between $T_H$ and $0$, so that  the vane rotates counter to the local precession but with the disk, the secondary rheological $\omega(T)$ curve has the same log-slope as for $T>0$ --- {\em the local precession rate is thus irrelevant for the resistance to flow}.
Only when the vane starts to move against the global disk
rotation for $T<T_H$, this slope changes.
Consistent with earlier work \cite{LosertPRL04}, moving with the 
fabric (i.e. co rotation) is tougher than moving against the fabric (counter rotation). We conclude that co and counter rotation should be defined with respect to the disk rotation, and not to the local precession rate.

\begin{figure*}[tb]
  \centering
    \includegraphics[width=2\columnwidth,clip,viewport = 0 260 570 500]{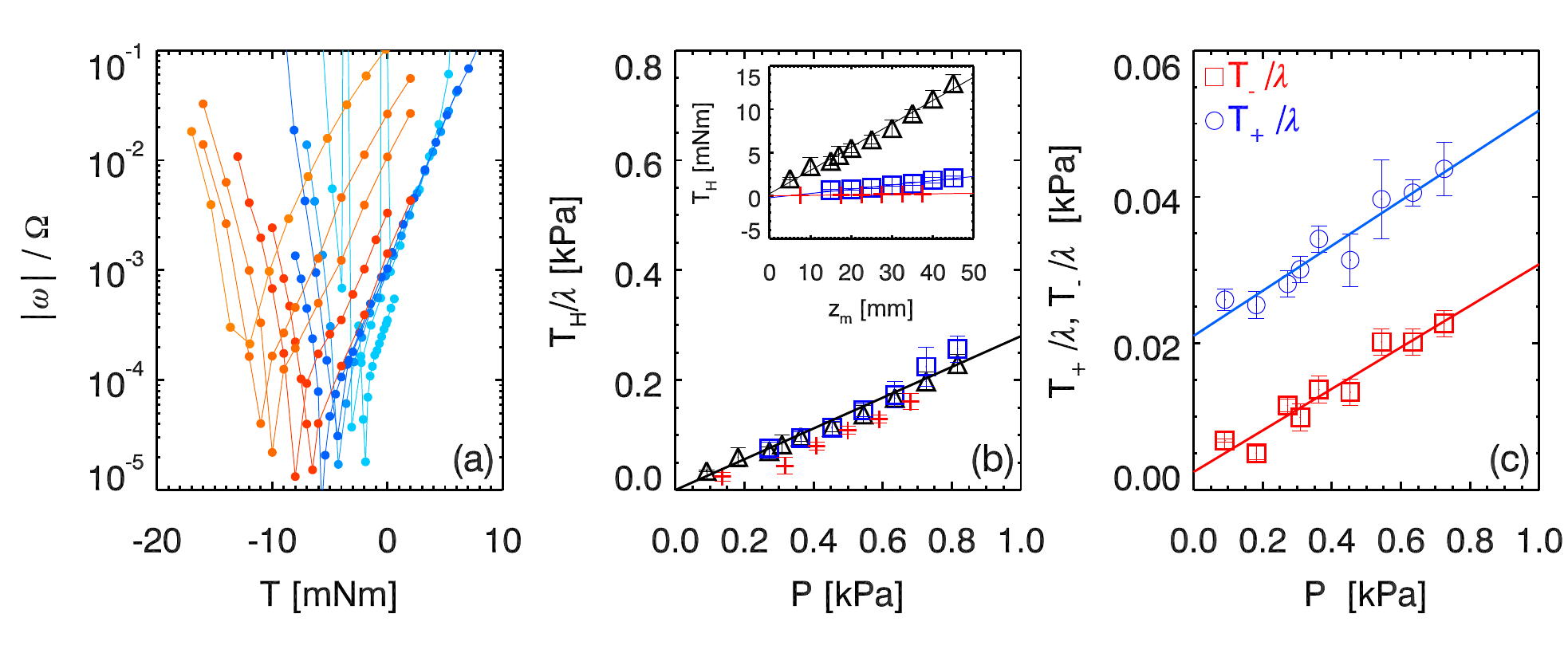}
\caption{(Color online) (a) $\omega/\Omega$ as a function of $T$ for depths $z_m=5,10,15,17,20,25,30,35,40$ and $45$ mm (right to left), for a $20\times20$ mm probe, $H=60$ mm, $\Omega=0.05$ rps. (b) Rescaled holding torques, $T_H/\lambda$ (Eq.~(3)), for probe sizes $5\times5$ mm (+), $10\times10$ mm ($\Box$) and
$20\times 20$ mm ($\triangle$). The slope of the straight line is 0.28, suggesting that $\mu_H \approx 0.28$. Inset: Holding torque as function of $z_m$. (c) The rescaled characteristic torques
$T_-/\lambda$ and $T_+/\lambda$ are linear
in the pressure $P$
(data shown for a $20\times20$ probe - the signal for smaller probes is too noisy to reliably determine $T_-$ and $T_+$).}\label{fig5}
\end{figure*}

To understand this, let us consider in more detail the flow in the 'mixed' regime where
$T_H<T<0$, and the vane rotates against the local precession but the with the disk rotation.
We know from experiments in
2D \cite{VejePRE99} and 3D \cite{LosertPRL04, BehringerNature05} that persistent shear leads to the buildup of an anistotropic fabric. This anisotropy is encoded in both
the contact and force structure in the material, and leads to a fragile structure that resists the current flow direction strongly, as revealed by experiments where reversing the flow direction leads to a transient in which the material has little resistance against flow, before
restructuring of the material with the opposite anisotropy leads to a steady state of high resistance.

In our experiments, we separate the generation of the main flow and the rheological probing.
Our data is consistent with a picture in which the rotation of the bottom disk creates an anisotropic fabric in the granular material, which ultimately leads to a difference in resistance to co flow and counter flow, and where the handedness of the anisotropy is
set by the main flow, and thus determined
by the sign of $\Omega$, even if the vane moves against the local precession rate.
So as long as $\omega/\Omega>0$, the vane rotates  against the anisotropy, even if $\omega< \omega_p$. In other words, the residual precession rate is irrelevant for the rheology experienced by the probe, although it does, of course, set the vane rotation for $T=0$.
By separating driving and probing, we thus can set up a steady state rotation of the probe with or against the anisotropic fabric, where the fabric is determined by the disk rotation only.

{\em Effect of depth ---} By changing the depth of the probe, $z_m$, we can probe the effect of pressure on the rheology, and probe how the holding torque $T_H$ and the characteristic torques $T_+$ and $T_-$ scale with pressure. In particular, we can probe whether these characteristic torques have a frictional nature --- if so, the torques, which can be translated into shear stresses (see Eq.~(3)), should be proportional to the pressure $P \propto z_m$.

In Fig.~\ref{fig5}a we show flow profiles
for a range of values of $z_m$. The deeper the probe, the more to the left the curves shift, and the  holding torque, where $\omega$ goes through zero, shows a systematic variation with probe depth $z_m$. In a frictional picture, one would
expect that $T_H$ would be proportional to the hydrostatic pressure $\sim z_m$, and taking the same corrections for the finite size of the probe into account as in our calculation of
the static friction coefficient, we would thus expect that $T_H/\lambda = \mu_H P$, where
$\lambda$ is defined as in Eq.~(\ref{lambdaeq}), and $\mu_H$ is an effective friction coefficient that determines the relation between holding torque and pressure.

As we shown in Fig.~\ref{fig5}b, the holding torque is indeed approximately linear in $z_m$, and also varies systematically with the probe size. When plotted as function of $P$, the data for $T_H/\lambda$ for all our probes collapse on a single straight curve, with slope $\mu_H \approx 0.28$.
The interpretation of this friction-like coefficient $\mu_H$ is not straightforward. We first note that it is roughly a factor two smaller than the static coefficient. It is also different from an ordinary dynamical friction coefficient, which would measure the resistance to flow of a granular medium, whereas here we measure the resistance to secondary flow in an {\em already flowing} granular medium. Notwithstanding the precise value of $\mu_H$, what is clear is that the holding torques are proportional to the pressure --- the rheology of secondary granular flows has a frictional character.

Motivated by this success, we plot in Fig.~\ref{fig5}c the values of $T_-/\lambda$ and $T_+/\lambda$ as a function of $P$. Surprising, we find that $T_-$ is also proportional to $P$ as $T_- /\lambda \approx 0.03 P$, whereas $T_+$ goes to a finite intercept for $P \rightarrow 0$, and $T_+/\lambda \approx 0.02+0.03P$. This implies that there is a significant qualitative difference between co and counter flow.

{\em Discussion and outlook ---} Our study shows that for secondary rheology, the anisotropy is a key parameter. Surprisingly, the difference between co and counter flow is not only quantitative, but qualitative:
In counter flow, the characteristic stresses $T_H$ and $T_-$ are both proportional to the pressure, whereas for co flow, the characteristic torque $T_+$ is not simply proportional to $P$. This surprising finding, and the difference between the scaling of $T_H$, $T_-$ and $T_+$ beg for further detailed (numerical) studies of 3D secondary granular rheologies.

Notwithstanding the complex variation of our characteristic torques with pressure, we note that our secondary
rheology can be written as $\omega(z)/\omega_p(z) = F(T/P,P)$ --- for counter flow, we get in good approximation that $\omega(z)/\omega_p(z) = \tilde{F}(T/P)$, so that the local flow is encoded in $\omega_p(z)$, whereas the secondary rheology is frictional. Our scaling for $T_-$ and $T_+$ suggests that for deeper layers, where the finite intercept of   $T_+$ becomes of less relative importance, we approach $T_- \approx T_+ \sim P$ --- so the effect of symmetry breaking diminishes with distance to the source of the main flow.

We note here that for the ``perpendicular'' probing, i.e., a measurement protocol not sensitive to
the flow direction, the hold torque is zero, and so in these cases it is more difficult to observe the frictional aspects of the secondary rheology.

Our secondary rheology has a rate which is clearly exponential in the shear stresses --- consistent with \cite{PouliquenPRL11} and \cite{AndreottiArxiv13}, but different from \cite{NicholPRL10}. We have no good explanation for this, but point out that the linear regime of \cite{NicholPRL10} was only observed for slow secondary flows, and it
is conceivable, that the observed linear regime simply corresponds to the overlap of two exponential functions.

Open questions for the future are to disentangle the role of pressure and distance to the source of the main flow, and models for the role of anisotropy in slow granular flows.

\acknowledgments
We acknowledge technical assistance of Jeroen Mesman. EW acknowledges funding by the Dutch funding agency FOM.


\begin{thebibliography}{99}
\bibitem {BehringerNature03}R. R. Hartley and B. P. Behringer \textit{Nature} \textbf{421} 928-931 (2003

\bibitem {DijksmanPRL11} J.A. Dijksman et. al. \textit{Phys. Rev. Lett} \textbf{107}, 108303 (2011)
\bibitem {DijksmanPRE10} J.A. Dijksman et. al.  \textit{Phys. Rev. E } \textbf{82}, 060301(R) (2010)


\bibitem {PouliquenAnnRevFlu08} Y. Forterre and O. Pouliquen, \textit{Ann, Rev. Fluid} \textbf{40} 1-24 (2008)
\bibitem{GDRMidi} $GDR$ Midi \textit{E. Phys. J. E} \textbf{14} 367-371 (2004)

\bibitem {PouliquenPRL11} K. A. Reddy, Y. Forterre and O. Pouliquen \textit{Phys. Rev. Lett} \textbf{106}, 108301 (2011)

\bibitem {NicholPRL10} K. Nichol et. al. \textit{Phys. Rev. Lett} \textbf{104}, 078302
(2010); K. Nichol and M van Hecke, Phys. Rev. E. {\bf85} 061309
(2012).

\bibitem {BocquetNature08} J. Goyon et. al. \textit{Nature}  \textbf{454}, 84-87 (2008)

\bibitem {KamrinPRL12} K. Kamrin and G. Koval \textit{Phys. Rev. Lett} \textbf{108} 178301 (2012)
\bibitem{AndreottiArxiv13} M. Bouzid et. al. \textit{ArXiv:cond-mat.soft} 1301.3308v2 (2013)
\bibitem {KamrinPNAS13} D. L. Henann and K. Kamrin \textit{Proc. Natl. Acad. Sci. USA} \textbf{110} 6730 (2013)
\bibitem {BocquetPRL09} L. Bocquet, A. Colin and A. Ajdari \textit{Phys. Rev. Lett} \textbf{103}, 036001 (2009)

\bibitem {KatgertEPL10} G. Katgert et. al. \textit{EPL} \textbf{90} 54002  (2010 )

\bibitem {FenisteinNature04} D. Fenistein and M. van Hecke, \textit{Nature} \textbf{425}, 256 (2003).
\bibitem {FenisteinPRL04} D. Fenistein, J.-W. van de Meent and M. van Hecke \textit{Phys. Rev. Lett} \textbf{92} 094301 (2004)
\bibitem {FenisteinPRL06} D. Fenistein, J.-W. van de Meent and M. van Hecke \textit{Phys. Rev. Lett} \textbf{96} 118001 (2006)
\bibitem {ChicagoPRL06} X. Cheng et. al., \textit{Phys. Rev. Lett.} \textbf{96} 038001 (2006)
\bibitem {joshuareview} J. Dijksman and M. van Hecke, \textit{Soft Matter} \textbf{6} 2901-2907 (2010)


\bibitem {LosertPRL04} M. Toiya, J. Stambaugh, and W. Losert  \textit{Phys. Rev. Lett.} \textbf{93} 088001 (2004)
\bibitem {BehringerNature05} T. S. Majmudar and R. P. Behringer \textit{Nature} \textbf{435} (23) 03805 2005


\bibitem {UngerPRL04} T. Unger, J. T\"or\"ok, J. Kert\'esz D. E.  Wolf \textit{Phys. Rev. Lett} \textbf{92} 214301 (2004)

\bibitem {VejePRE99} C.T. Veje, Daniel W. Howell and R. P. Behringer \textit{Phys. Rev. E} \textbf{59} 739 (1999)






\end{thebibliography}
\end{document}